\documentstyle[epsfig,longtable]{aipproc}

\begin{document}
\def \cosw{\cos \chi_w^t}
\def \sinw{\sin \chi_w^t}
\def \ce{\cos \chi_e^w}
\def \se{\sin \chi_e^w}
\def \sse{\sin^2 \chi_e^w}
\def \pew{\Phi}
\def \chie{\chi_e^w}
\def \chiw{\chi_w^t}
\def \sw{\sin \theta_W}
\def \ssqw{\sin^2 \theta_W}
\def \cw{\cos \theta_W}
\def \thetad{\theta^{\dagger}}
\def \thetas{\theta^*}
\def \sangle{\xi}
\def \ds{\frac{d \sigma}{d \cos \theta^*}}
\def \norm{ \Bigl( ~\frac{3 \pi \alpha^2}{2 s}\beta ~\Bigr) ~}
\def \norme{\Bigl( ~\frac{3 \pi \alpha^2}{8 s}\beta ~\Bigr) ~}
\def \norms{\Bigl( ~\frac{ \pi \alpha_s^2}{9 s}\beta ~\Bigr) ~}
\def \beq{\begin{equation}}
\def \eeq{\end{equation}}
\def \beqa{\begin{eqnarray}}
\def \eeqa{\end{eqnarray}}
\def \amp{{\cal M}}
\def \ttbar{$t \bar t$}
\def \qqbar{$q \bar q$}
\def \antibeamline{``anti-beamline''}
\def \beamline{Beamline}
\def \UP{{\rm R}}
\def \DOWN{{\rm L}}
\def \up{\uparrow}
\def \down{\downarrow}
\def \half{\hbox{$1\over2$}}
\def \qbar{\bar q}
\def \tbar{\bar t}
\def \ebar{\bar e}
\def \opt{Off-diagonal}
\def \epm{$e^+e^-$}
\def \eL{$e^-_L~e^+_R$}
\def \eR{$e^-_R~e^+_L$}\vspace*{-1in}

\begin{flushright}
\small{FERMILAB-Conf-98/056-T \\
hep-ph/9802279\\
February 98}
\end{flushright}

\def \UD{$t_{\up} ~{\bar t}_{\down}$ }
\def \DU{$t_{\down} ~{\bar t}_{\up}$ }
\def \UU{$t_{\up} ~{\bar t}_{\up}$ }
\def \DD{$t_{\down} ~{\bar t}_{\down}$ }
\def	\nn		{\nonumber}
\def	\=		{\;=\;} 
\def	\ret		{\\[\eqskip]}
\def	\to		{\rightarrow }

\title{Top Quark Physics\\ at a Polarized Muon Collider
\footnote{
Presented at the workshop on ``Physics at the First Muon Collider,''
November 6-7, 1997 at Fermilab, Batavia, IL and at the 
``Fourth International Conference on the Physics Potential and Development
of Muon Collider,'' Dec 10-12, 1997, San Francisco, CA.
}}

\author{Stephen Parke}
\vspace*{-0.5in}
\address{Theoretical Physics Department\\
Fermi National Accelerator Laboratory
\\
Batavia, IL 60510\\ USA \\ e-mail: parke@fnal.gov}
\maketitle

\begin{abstract}
Top quark pair production is presented at a polarized Muon
Collider above the threshold region.
The off-diagonal spin basis is the natural basis for this discussion
as the top quark pairs are produced in an essentially unique spin
configuration for 100\% polarization.
Modest polarization, say 30\%, can lead to 90\% of all top quark 
pair events being in one spin configuration.
This will lead to sensitive tests on anomalous top quark couplings.
\end{abstract}

\begin{figure}[b!] 
\centerline{\epsfig{file=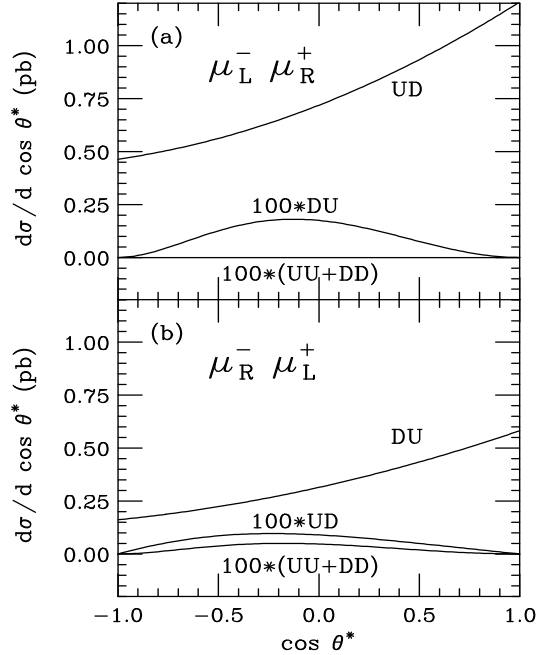,height=3.4in,width=2.75in}}
\vspace{10pt}
\caption{The spin configurations using the off-diagonal basis for both
$\mu^+_L~\mu^-_R$ and $\mu^+_R~\mu^-_L$ for
$\sqrt{s}=400~GeV$.
Note that the sub-leading configurations have been amplified 
by a factor of 100 in these figures.}
\label{fig1}
\end{figure}

Recently Parke and Shadmi \cite{PS} have shown that at a 100\% polarized 
lepton collider that the top and anti-top quark pairs are produced in
essentially a unique spin configuration.
This spin basis has been called the ``off-diagonal'' basis and it interpolates
between the beam direction at threshold and the top quark direction,
i.e. helicity, far above threshold. 
The differential cross section using this basis is given by
\beqa
& & \ds ~(\mu^-_L~\mu^+_R \rightarrow t_{\up}~\tbar_{\up} 
~or~  t_{\down}~\tbar_{\down})   =  0 \ , \nn \\[0.15in]
& & \ds ~(\mu^-_L~\mu^+_R \rightarrow t_{\up}~\tbar_{\down} 
 ~or~  t_{\down}~\tbar_{\up})  =  
\norme \nn \\[0.15in] & & 
\quad \Biggl[ ~f_{LL}(1 + \beta \cos \thetas) 
+ f_{LR}(1 - \beta \cos \thetas) 
\nn \\[0.15in] & &
\quad \quad \pm  \sqrt{ 
(f_{LL}(1 + \beta \cos \thetas)
 -  f_{LR}(1 - \beta \cos \thetas))^2 
+4f_{LL}f_{LR}(1-\beta^2)~}~ \Biggr] ^2 \  
\label{Optimal}
\eeqa
where the $f_{IJ}$'s are the sum of the photon 
and Z-boson couplings corrected for the difference in the propogators.
Details of this basis can be found in reference \cite{PS}.
Figure \ref{fig1} is the spin components for top quark pair production
in the off-diagonal basis for both LR and RL incoming lepton helicities for
a $\sqrt{s}~=~400~GeV$ collider.
The sub-leading terms have been amplified by a factor of 100 so it is clear
that the dominant configuration makes up more than 99\% of the total cross
section. 

\begin{figure}[b!] 
\centerline{\epsfig{file=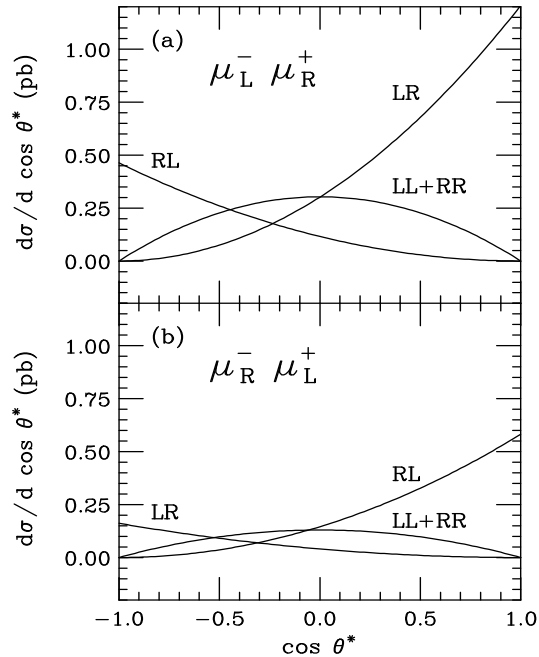,height=3.4in,width=2.75in}}
\vspace{10pt}
\caption{The spin configurations using the helicity basis for a 
$\sqrt{s}=400~GeV$ collider.
}
\label{fig2}
\end{figure}

In the helicity basis the differential cross section is \cite{PSch}
\beqa
\ds ~(\mu^-_L~\mu^+_R \rightarrow t_L~\tbar_L ~or~ t_R~\tbar_R) & = &
 \norme ~(1-\beta^2) \sin^2 \thetas~|f_{LL}+f_{LR}|^2  \ , \nn \\[0.1in]
\ds ~(e^-_L~e^+_R \rightarrow t_R~\tbar_L
~or~ t_L~\tbar_R) & = &
\norme ~(1\mp \cos \thetas)^2 \nn \\
& & \quad \quad \times |f_{LL}(1 \mp \beta) + f_{LR}(1 \pm \beta)|^2 \ .
\label{Helicity}
\eeqa
Figure \ref{fig2} is the corresponding plot for the helicity
basis. 
Here the dominant spin configuration is less than 60\% of the total.

\begin{figure}[b!] 
\centerline{\epsfig{file=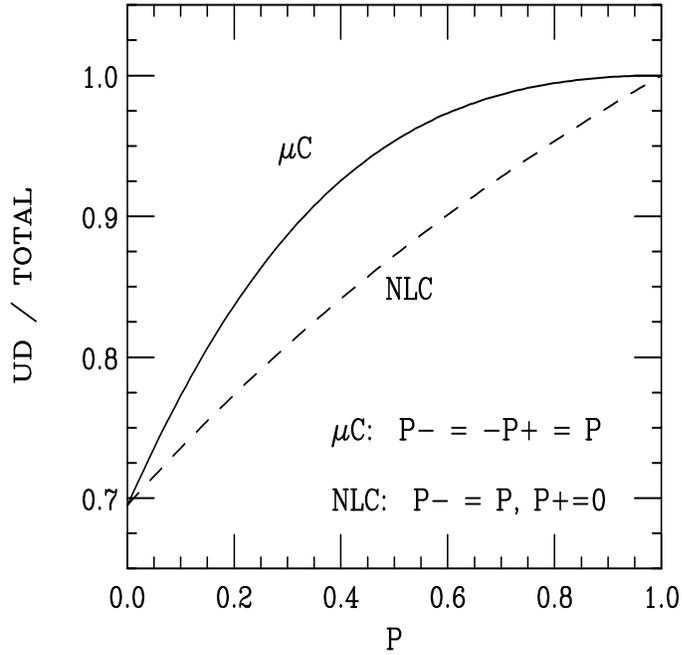,height=3.4in,width=3.5in}}
\vspace{10pt}
\caption{Fraction of the total cross section in the off-diagonal basis' 
Up-Down spin configuration as a function of the polarization. 
Both beams are assumed to be polarized for the Muon Collider ($\mu$C)
but only one beam for the NLC.
}
\label{fig3}
\end{figure}

In figure \ref{fig3} I have plotted the dominant spin component's 
fraction of the total as function of the polarization of the beams,
\begin{equation}
{{(1+P^-)(1-P^+)\sigma_{LR \rightarrow UD} ~+~
 (1-P^-)(1+P^+)\sigma_{RL \rightarrow UD} }
\over
{(1+P^-)(1-P^+)\sigma_{LR}^{tot} ~+~
 (1-P^-)(1+P^+)\sigma_{RL}^{tot}  }}
\end{equation}
for the off-diagonal basis. 
Here $P \equiv (N_L-N_R)/(N_L+N_R)$ for 
the $\mu^+$ and $\mu^-$ beams. 
Two different machines are included, the
Muon Collider and the NLC. 
The Muon Collider is assumed to have equal
but opposite polarization for the $\mu^+$ and $\mu^-$ beams
whereas the NLC is an electron-positron collider with only the electron beam
polarized. 
From these curves a modest amount of polarization, say 30\%, at
a Muon Collider can make the dominant spin configuration close to 90\%
of the total. Whereas at an electron-positron machine one requires 55\%
polarization to achieve the same goal.

\begin{figure}[b!] 
\centerline{\epsfig{file=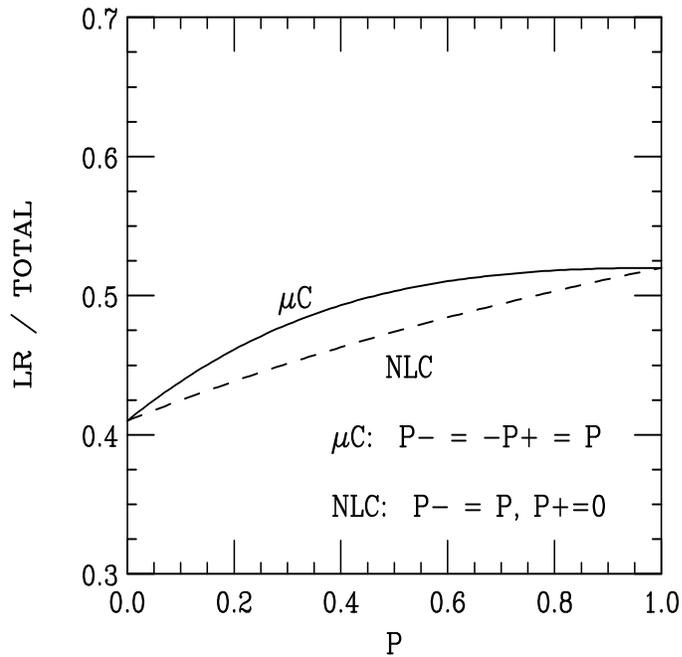,height=3.4in,width=3.5in}}
\vspace{10pt}
\caption{Same as Fig. 3 but the helicity basis is used.}
\label{fig4}
\end{figure}

\begin{figure}[b!] 
\centerline{\epsfig{file=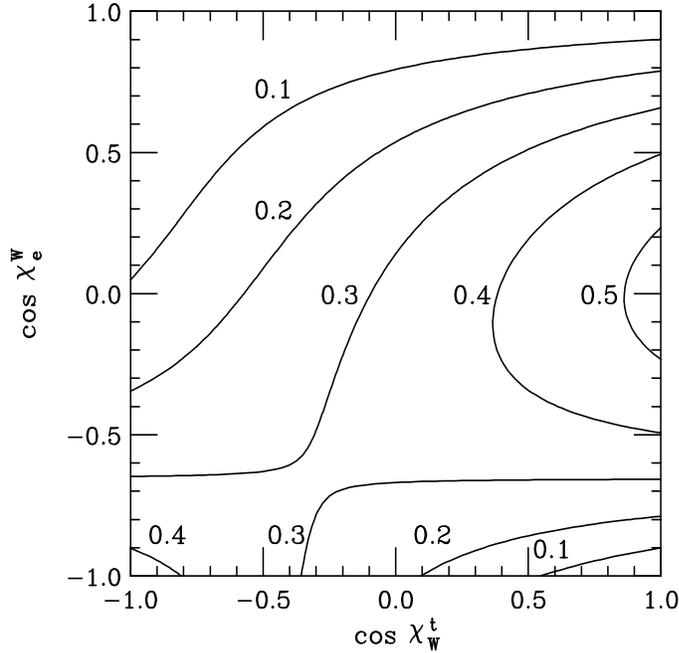,height=3.4in,width=3.5in}}
\vspace{10pt}
\caption{Correlations between the W-boson and the top spin 
direction in the top rest frame ($\cos \chi^t_W$) 
and the positron and the minus b-quark
direction in the W-boson rest frame ($\cos \chi^W_e$).}
\label{gamma}
\end{figure}

\begin{figure}[b!] 
\centerline{\epsfig{file=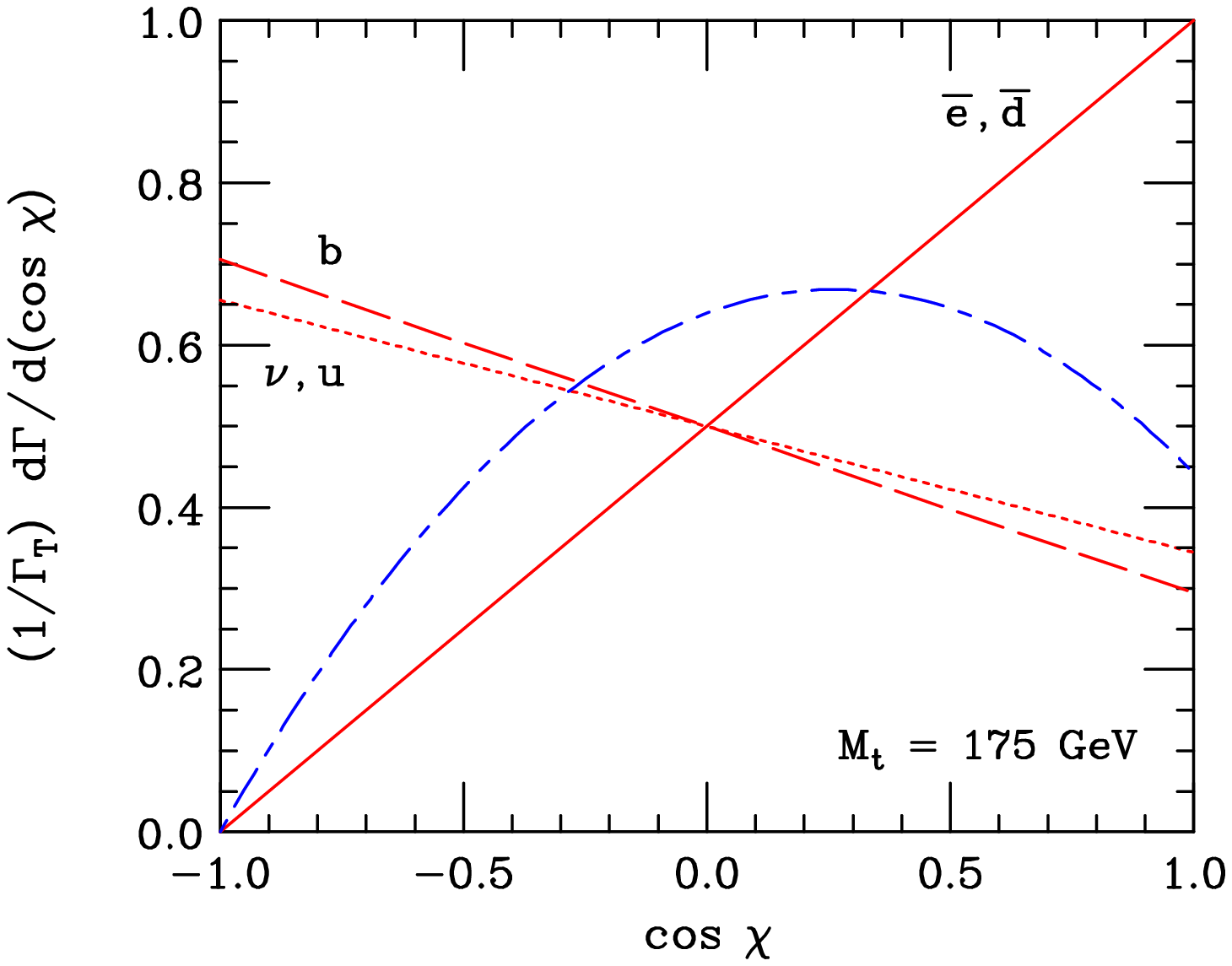,height=3.0in,width=3.5in}}
\vspace{10pt}
\caption{The straight lines are the correlations between the top quark 
decay products and the top quark spin direction in the rest frame of the 
top ($\cos \chi^t_i$).
Whereas the curved line is the correlation between the b-quark and the positron
or d-quark in the rest frame of the W-boson ($-\cos \chi^W_e$).}
\label{alpha}
\end{figure}

Figure \ref{fig4} is the corresponding plot for the helicity configuration.
If one uses this spin basis the dominant spin configuration ranges from 
41\% to 52\% of the total. Clearly polarization is not as important here
without further cuts.

Since the top-quark pairs are produced in a nearly unique spin configuration,
and the electroweak decay products of polarized top-quarks are 
strongly correlated to the spin axis, 
the top-quark events at  $\mu^+~\mu^-$ collider have a
very distinctive topology. 
Deviations from this topology would signal anomalous couplings.
In the Standard Model, the predominant decay mode of the top-quark is 
$t \rightarrow b W^+$, with the $W^+$ 
decaying either hadronically or leptonically. 
For definiteness we consider here the decay 
$t \rightarrow b W^+\rightarrow b e^+ \nu$. 
The differential decay width of a polarized top-quark 
 depends non-trivially on three angles. 
The first is the angle, $\chiw$, between the top-quark spin and 
the  direction of motion of the $W$-boson in the top-quark rest-frame.
Next is the angle between the direction of 
motion of the $b$-quark and the positron in the W-boson rest-frame. 
We call this angle $\pi - \chie$. 
Finally, in the top-quark rest-frame, we have the azimuthal angle, $\pew$,
between the positron direction of motion and the top-quark spin 
around the direction of motion of the $W$-boson.

The differential polarized top-quark decay distribution in terms of 
these three angles is given by
\beqa
\frac{1}{\Gamma_T} ~{d^3 ~\Gamma \over d \cosw ~d \ce ~d \pew} & = &
\frac{3}{8 ~(m_t^2+2m_W^2)}\nn \\[0.1in]
~\Bigl[m_t^2(1+\cosw)\sse  & + & m_W^2(1-\cosw)(1-\ce)^2 \nn \\[0.1in] 
& + & 2 m_t m_W (1-\ce)\se \sinw \cos \pew  \Bigr] \ ,
\label{diffgamma}
\eeqa
where $m_t$ is the top-quark mass, $m_W$ is the $W$ mass,
and $\Gamma_T$ is the total decay width
(we neglect the  $b$-quark mass).
The first and second terms in~(\ref{diffgamma}) give the contributions 
of longitudinal and transverse $W$-bosons respectively. 
The interference term, given by the third 
term in~(\ref{diffgamma}), does not contribute to the total width, 
but its effects on the angular distribution of 
the top-quark decay products are sizable.
Fig.~\ref{gamma} shows 
contour plots 
of the differential angular decay distribution
in the $\chie-\chiw$ plane
after integrating over the azimuthal angle 
$\pew$.
The peak at the center of the right hand side of this figure is due to the
longitudinal $W$-bosons whereas the peak at the bottom left hand corner is
caused by the transverse $W$-bosons.

There are also significant correlations 
of the angle between the top-quark spin  
and the momentum of the $i$-th decay product, $\chi^t_i$,
measured in the top-quark rest-frame, see figure \ref{alpha}.
The differential decay rate of the top-quark is given by
\beqa
\frac{1}{\Gamma_T} ~{d ~\Gamma \over d \cos \chi^t_i} & = &
\frac{1}{2}
~\Bigl[1 + \alpha_i \cos \chi^t_i \Bigr] \ ,
\eeqa
where $\alpha_b = -\alpha_W = -0.41$, $\alpha_{\nu} = -0.31$ 
and $\alpha_{e^+} = 1$,
for $m_t = 175$ GeV,
see ref.~\cite{Jezabek}.
The interference between the longitudinal and transverse $W$-bosons
is very importnat in determining these correlations.
Note, the positron is more highly correlated with the spin of 
the top quark than its parent the $W$-boson!

Given that we know the spin configuration of the top quark pairs 
and the correlations of the top quark decay products 
there are many correlations studies that can be performed 
in top quark pair product at a lepton collider looking 
for anomalous couplings of the top quarks.
These studies have been performed for the helicity basis \cite{BS}
but need to be redone using the superior off-diagonal spin basis.

QCD effects modify this picture in only a minor fashion. 
The reason being that soft gluons cannot flip the spin of the heavy
top quarks. Detailed studies of the effects of one loop calculations
show that the dominant spin configuration is still more than 99\% of the
total even when QCD corrections are included \cite{KNP}.

In conclusion top quark pairs above the threshold region at Muon Colliders
are a great place to search for anomalous couplings of the top quark.
For $\sqrt{s} ~<~ 1 ~TeV$ the off-diagonal basis is superior to the helicity
basis in describing the events in the simplest possible terms.
Polarization of the incoming beams enhances this effect.
Detailed studies of the one loop QCD corrections have been recently completed,
showing no qualitative difference than the tree level analysis.
An extensive analysis of the sensitivity to 
anomalous couplings of the top quark,
using the off-diagonal basis, is now needed.

\section*{ACKNOWLEDGMENTS}
Special thanks to the local organizers of this conference. 
Fermi National Accelerator Laboratory is operated by 
the Universities Research Association, Inc., under contract
DE-AC02-76CHO3000 with the United States of America Department of Energy.

\end{document}